\documentclass[11pt]{article}
\usepackage[preprint]{acl}
\usepackage{times}
\usepackage{latexsym}
\usepackage[T1]{fontenc}
\usepackage[utf8]{inputenc}
\usepackage{microtype}
\usepackage{graphicx}
\usepackage{booktabs}
\usepackage{multicol}
\newsavebox{\tablecontentbox}
\newcommand{\fitwidetable}[1]{%
  \sbox{\tablecontentbox}{#1}%
  \ifdim\wd\tablecontentbox>\textwidth
    \resizebox{\textwidth}{!}{\usebox{\tablecontentbox}}%
  \else
    \usebox{\tablecontentbox}%
  \fi
}
\usepackage{amsmath,amssymb}
\usepackage{url}
\providecommand{\tightlist}{\setlength{\itemsep}{0pt}\setlength{\parskip}{0pt}}
\title{RSE-TTS: Compact Text-to-Speech through Shared Text-Aware Reference
Encoding and Generation}
\author{\textbf{Ming Ma\quad Mingjie Chen\quad Yuanzhong Zheng\quad Hengyao Bao\quad Hailin Yu} \\[3pt]
\textbf{Shijing Yu\quad Hao Qiu\quad Gongfan Cao\quad Xiao Li\quad Xuemin Zhao\quad Youhai Jiang} \\[5pt]
VoxNexus \\[2pt]
{\small\normalfont \{ming.ma, mingjie.chen, yuanzhong.zheng, hengyao.bao, hailin.yu, shijing.yu\}@voxnexus.ai} \\[1pt]
{\small\normalfont \{hao.qiu, gongfan.cao, xiao.li, xuemin.zhao, youhai.jiang\}@voxnexus.ai}}
\begin{document}
\maketitle
\begin{abstract}
Can a speech generator also serve as an effective voice encoder? We show
that a pretrained autoregressive text-to-speech (TTS) backbone can be
repurposed for reference compression, improving speaker similarity in
subsequent synthesis. Jointly encoding reference speech and its
transcript further strengthens this capability. Controlled ablations and
activation interventions trace the text benefit through intermediate
acoustic representations into the compressed vectors that condition
generation. Building on these findings, we introduce RSE-TTS, whose
Reconstructive Speaker Encoder (RSE) shares the generation backbone and
compresses paired reference text and speech into eight continuous
vectors. Sharing reduces encoder--generator parameters by 48\% relative
to independent pretrained encoding. Large-scale English training yields
a 118M-parameter generator that improves both word error rate and
speaker similarity over the corresponding official MOSS-TTS-Nano cloning
modes on all three English benchmarks. Combined with acoustic
continuation, RSE-TTS approaches the speaker similarity of substantially
larger systems while achieving lower WER than the evaluated CosyVoice 3
and MOSS-TTS-Local models. Audio samples are available at
\url{https://voxnexusai.github.io/RSE-TTS/}.
\end{abstract}
\begin{figure*}[t]
\centering
\includegraphics[width=\textwidth]{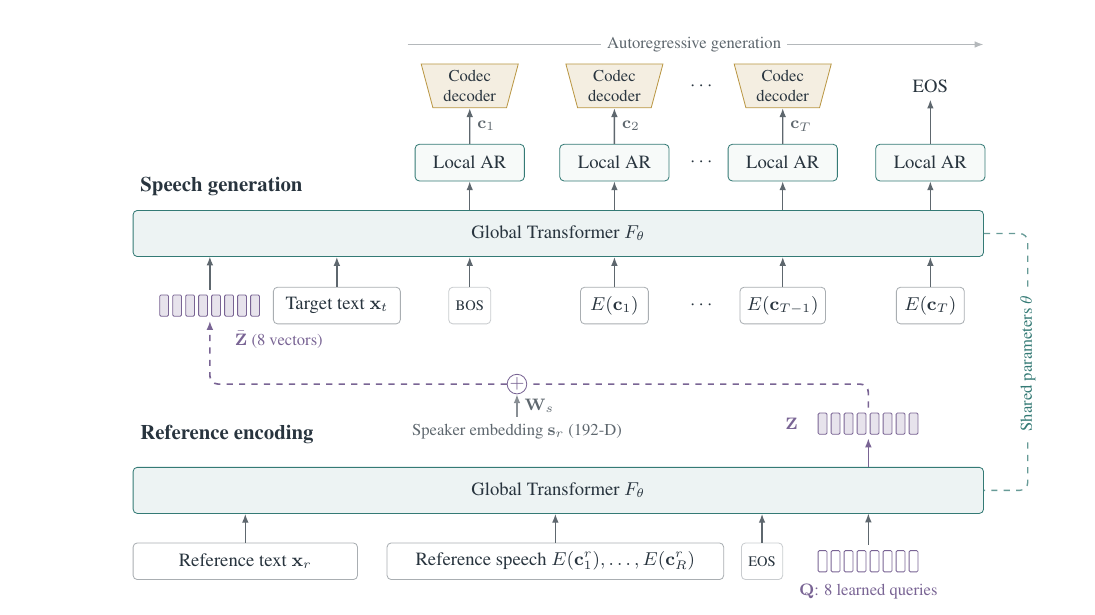}
\caption{RSE-TTS uses the same Global transformer to encode reference text and
speech and to generate a new utterance. Eight learned queries produce
\(\mathbf{Z}\), which is combined with a projected speaker embedding to
form \(\bar{\mathbf{Z}}\). During generation, Global consumes frame
embeddings \(E(\mathbf{c}_i)\), while Local produces discrete codec
frames \(\mathbf{c}_i\). Repeated Local and decoder blocks illustrate
successive generation steps, not separate parameters.}
\end{figure*}

\section{Introduction}\label{introduction}

Scaling model capacity and training data has driven substantial advances
in text-to-speech synthesis, improving pronunciation accuracy, speaker
similarity, and expressive naturalness. These gains have been
demonstrated across a broad range of speech generation architectures
\citep{hu2026qwen3tts, jia2025ditar, zhou2025voxcpm, wang2024maskgct, chen2024f5tts, du2025cosyvoice3}.
However, bringing these capabilities to compact models remains
challenging: a small generator must accurately realize new linguistic
content while preserving the identity and delivery of a reference
speaker. This makes effective use of reference speech a central design
consideration for compact TTS.

Acoustic continuation (prompting) provides a direct way to preserve rich
voice information by treating reference speech as part of the generation
history \citep{wang2023valle}. This access, however, couples target
synthesis to the reference utterance. Longer prompts consume more
generation context, while recording artifacts, atypical pauses, and
mismatches between reference speech and its transcript can complicate
synthesis \citep{zhang2025minimax}. The acoustic history therefore
carries both useful voice cues and reference-specific content and
conditions that need not be reproduced.

A reference encoder separates reference processing from target
generation by summarizing the recording into a fixed-size condition
\citep{zhang2025minimax}. This interface supplies explicit voice cues
from the onset of synthesis without retaining the full reference
history. Its central challenge is preserving sufficient speaker identity
and expressive detail through compression \citep{zhou2022cdfse}. To
address this challenge, we investigate whether a pretrained TTS
generator's own text--speech modeling capabilities can be reused to
construct richer reference representations.

We find that pretrained generation capabilities transfer to reference
compression: initializing an independent encoder from the TTS backbone
substantially improves generated speaker similarity over random
initialization. Paired reference text further improves both independent
and shared encoders. To understand this benefit, we disrupt text--speech
correspondence and intervene on internal states. Restoring intermediate
reference-audio or later query representations recovers generated
speaker similarity, linking reference understanding to voice
reconstruction.

We turn this capability into a compact system, \textbf{RSE-TTS}. Its
\textbf{Reconstructive Speaker Encoder (RSE)} jointly compresses
reference text and speech into eight vectors. Encoding and generation
share one backbone, avoiding a second Transformer and reducing their
combined parameter count by 48\%. We instantiate the design in
MOSS-TTS-Nano \citep{openmoss2026nano} and train it through target
speech prediction.

Training on a large-scale in-house English speech corpus yields a
118M-parameter generator that improves both intelligibility and speaker
similarity over official MOSS-TTS-Nano across three English benchmarks.
Acoustic continuation further improves similarity, bringing the compact
system closer to substantially larger models. Our contributions are:

\begin{itemize}
\tightlist
\item
  \textbf{Generative capabilities for voice compression.} We demonstrate
  transfer from pretrained AR TTS to reference encoding, with paired
  text further improving speaker similarity in both independent and
  shared encoders.
\item
  \textbf{Mechanistic understanding of text-aware reference encoding.}
  Transcript perturbations and activation interventions reveal how
  reference text affects acoustic representations and compressed query
  states that support voice reconstruction.
\item
  \textbf{Parameter-efficient reference encoding and synthesis.} RSE-TTS
  shares the encoding and generation backbone, reducing their combined
  parameter count by 48\% relative to independent encoding. Large-scale
  training demonstrates the design's effectiveness in a compact TTS
  system, with acoustic continuation providing a further improvement in
  speaker similarity.
\end{itemize}

\section{Related Work}\label{related-work}

\textbf{Reference conditioning for zero-shot TTS.} Speaker-verification
encoders transfer identity representations to TTS
\citep{jia2018speaker}, while MiniMax-Speech, IndexTTS2, and XTTS learn
synthesis-oriented reference conditioners
\citep{zhang2025minimax, zhou2026indextts2, casanova2024xtts}. VALL-E
retains reference speech as acoustic context \citep{wang2023valle};
Koel-TTS compares speaker embeddings, decoder-context conditioning, and
a separate context encoder \citep{hussain2025koel}. RSE reuses the
generation backbone in a separate text-aware encoding pass, sharing its
learned capabilities while compressing the reference into eight vectors
for target generation.

\textbf{Text-aware voice and style representations.} Prosody encoders
transfer acoustic expression \citep{skerryryan2018prosody}, while global
style tokens summarize expressive variation \citep{wang2018gst}.
Content-dependent speaker embeddings select reference features relevant
to target phonemes \citep{zhou2022cdfse}, and Mega-TTS 2 models timbre
and prosody with longer and multiple references
\citep{jiang2024megatts2}. MMCE-Qformer uses separate Q-Formers to
summarize historical text and speech, with the current synthesis text
guiding acoustic-context extraction for context-aware TTS
\citep{xue2024mmce}. RSE jointly encodes a reference recording and its
own transcript with a pretrained TTS backbone, using its learned
text--speech relationships to construct a target-independent voice
condition. Continuous context compression has also been explored in text
language models \citep{chevalier2023autocompressors, ge2024icae}. RSE
applies this idea to paired text and speech, preserving a speaker's
voice across changes in linguistic content.

\section{Reconstructive Speaker
Encoder}\label{reconstructive-speaker-encoder}

\subsection{Shared Global--Local speech
model}\label{shared-globallocal-speech-model}

We build on MOSS-TTS-Nano, a compact autoregressive codec TTS model with
a Global--Local transformer architecture \citep{openmoss2026nano}. The
Global transformer models the sequence of text and acoustic frames using
12 layers with a hidden dimension of 768, while the single-layer Local
transformer autoregressively generates the codebook indices within each
frame. The frozen codec represents speech at 12.5 frames per second
using 16 codebooks and reconstructs audio at 48 kHz. Codec
reconstruction preserves speaker similarity well (Appendix H),
motivating our focus on reference encoding and speech generation.

Let \((\mathbf{a}_r,\mathbf{x}_r)\) denote a reference recording and its
transcript, and let \((\mathbf{a}_t,\mathbf{x}_t)\) denote a target
utterance from the same speaker.\footnote{Bold lowercase and uppercase
  letters denote vectors and matrices, respectively.} The codec maps a
recording to a sequence of frame-level codebook tuples,
\(C(\mathbf{a})=(\mathbf{c}_1,\ldots,\mathbf{c}_T)\). Each
\(\mathbf{c}_i\) contains the 16 codec indices at frame \(i\), and
\(E(\mathbf{c}_i)\) denotes its continuous input embedding. The
embeddings of the codec tokens within each frame are summed to form a
single frame-level input to Global. Let \(F_\theta\) denote the shared
Global transformer and
\(\mathbf{Q}=[\mathbf{q}_1,\ldots,\mathbf{q}_8]\in\mathbb R^{d\times8}\)
the matrix of eight learned query embeddings, with \(d=768\). The
reference recording yields \(R\) frames, embedded as
\(E(\mathbf{c}^r_1),\ldots,E(\mathbf{c}^r_R)\).

\subsection{Encoding the reference}\label{encoding-the-reference}

The reference input sequence \(\mathbf{S}_r\) consists, in order, of the
reference text embeddings, reference speech embeddings, an end-of-speech
marker (EOS), and the eight learned queries. Writing the embedded
sequence as columns of \(\mathbf{S}_r\in\mathbb R^{d\times N_r}\), where
\(N_r\) is its length, we obtain hidden states
\(\mathbf{H}_r\in\mathbb R^{d\times N_r}\) with the shared backbone:

\[
\mathbf{H}_r=F_\theta(\mathbf{S}_r).
\]

Each query attends to the preceding reference text and speech under the
causal mask. All eight query outputs are computed in one transformer
pass. Let \(\mathbf{h}^r_{q,j}\) denote the output at query position
\(j\). Layer normalization and a learned linear bridge produce the
compressed vectors:

\[
\mathbf{z}_j=B_\phi\bigl(\operatorname{LN}(\mathbf{h}^r_{q,j})\bigr),\qquad j=1,\ldots,8.
\]

Here,
\(\mathbf{Z}=[\mathbf{z}_1,\ldots,\mathbf{z}_8]\in\mathbb R^{d\times8}\)
stacks the compressed reference vectors as columns. We retain the
existing 192-dimensional speaker-conditioning path, using a frozen CAM++
extractor \citep{wang2023campplus}. If
\(\mathbf{s}_r\in\mathbb R^{192}\) is the reference speaker embedding,
its normalized projection is added to each vector:

\[
\bar{\mathbf{z}}_j=\mathbf{z}_j+\mathbf{W}_s\frac{\mathbf{s}_r}{\|\mathbf{s}_r\|_2}.
\]

Here, \(\mathbf{W}_s\in\mathbb R^{d\times192}\). The resulting condition
is
\(\bar{\mathbf{Z}}=[\bar{\mathbf{z}}_1,\ldots,\bar{\mathbf{z}}_8]\in\mathbb R^{d\times8}\).
We study the information provided by \(\mathbf{Z}\) beyond this
auxiliary speaker condition in Section 5.

\subsection{Reconstructing same-speaker
speech}\label{reconstructing-same-speaker-speech}

The target branch uses the same Global transformer \(F_\theta\). Its
input begins with \(\bar{\mathbf{Z}}\), followed by the complete target
text and a speech-start marker, labeled BOS in Figure 1. Target
generation receives the compressed condition rather than the full
reference history.

At each generation step, the model predicts whether to continue or stop.
When continuing, the Local transformer generates the next frame's codec
tokens autoregressively, conditioned on the Global output. The resulting
frame is embedded and fed back to Global for the next step. Generation
ends when the model predicts EOS, and the codec decoder converts the
generated frames into speech.

\subsection{Training objective and computational
implications}\label{training-objective-and-computational-implications}

We train with teacher forcing using the autoregressive cross-entropy of
the target speech, including its end-of-speech prediction. This single
objective jointly trains reference encoding and target generation:
gradients flow from the target predictions through \(\bar{\mathbf{Z}}\)
into the query readout and shared backbone. Loss weights are specified
in Appendix A.

The same-speaker pairing requires the compressed representation to
support a different utterance from the reference. We share the backbone
to reduce the model's parameter footprint, avoiding a second full
transformer for reference encoding. Longer references increase the work
of the reference pass, while target generation always receives the same
eight-vector interface.

\section{Experimental Setup}\label{experimental-setup}

\subsection{Training data and model
variants}\label{training-data-and-model-variants}

We construct same-speaker utterance pairs from an in-house English
speech corpus, selecting target utterances based on acoustic quality and
transcription accuracy. Reference utterances are sampled from the same
speaker. Complete reference and target sequences are retained. We apply
IPA augmentation to both reference and target transcripts by
probabilistically replacing unambiguous words with their phonetic
transcriptions.

We train five variants for one epoch on the same \textbf{20k-hour
subset}, separating initialization, reference text, and parameter
sharing. All generators start from the same English TTS checkpoint,
pretrained with a small convolutional speaker encoder. We replace that
encoder with a Transformer matching Global:

\begin{enumerate}
\def\labelenumi{\arabic{enumi}.}
\tightlist
\item
  \textbf{Independent audio (random):} a randomly initialized reference
  backbone receives speech alone and trains independently of the
  generator.
\item
  \textbf{Independent audio (pretrained):} the reference backbone copies
  the pretrained generator's weights before training, then updates
  independently.
\item
  \textbf{Independent text + audio:} the pretrained independent encoder
  additionally receives the paired reference transcript.
\item
  \textbf{Shared audio:} the reference encoder uses the generator's
  parameters and receives speech alone.
\item
  \textbf{Shared text + audio (RSE):} the shared backbone receives the
  paired transcript and recording.
\end{enumerate}

All variants use the same data order seed, learning-rate schedule, eight
queries, auxiliary speaker condition, and target-only objective.

Beyond the controlled ablations, we train the full RSE-TTS system on a
large-scale in-house English speech corpus and compare it with existing
TTS systems. Controlled-ablation training and inference settings are
provided in Appendix A.

\begin{table*}[t]
\centering\small
\fitwidetable{%
\begin{tabular}{lrrrr}
\toprule
Reference interface & Params (M) & \begin{tabular}{@{}c@{}}\textbf{Seed-TTS}\\{\footnotesize WER (\%) $\downarrow$ / SIM $\uparrow$}\end{tabular} & \begin{tabular}{@{}c@{}}\textbf{CV3-eval}\\{\footnotesize WER (\%) $\downarrow$ / SIM $\uparrow$}\end{tabular} & \begin{tabular}{@{}c@{}}\textbf{MiniMax}\\{\footnotesize WER (\%) $\downarrow$ / SIM $\uparrow$}\end{tabular} \\
\midrule
Independent audio (random) & 228.5 & 1.39 / 0.400 & \textbf{2.44} / 0.342 & 0.75 / 0.391 \\
Independent audio (pretrained) & 228.5 & 1.46 / 0.530 & 2.63 / 0.469 & 0.91 / 0.592 \\
Independent text + audio & 228.5 & 1.41 / \textbf{0.564} & 2.50 / \textbf{0.507} & 0.75 / \textbf{0.651} \\
Shared audio & 118.2 & 1.35 / 0.498 & 2.46 / 0.435 & 0.80 / 0.553 \\
Shared text + audio (RSE) & 118.2 & \textbf{1.33} / 0.552 & 2.63 / 0.499 & \textbf{0.64} / 0.615 \\
\bottomrule
\end{tabular}}
\caption{Five encoder variants trained on the same 20k hours. Independent
encoders are pretrained unless labeled random. Params counts unique
trainable encoding-and-generation parameters, excluding common frozen
modules. WER is corpus-level text-normalized WER (\%); bold marks the
best value. Prompts and scoring are matched.}
\end{table*}

\subsection{Evaluation}\label{evaluation}

We evaluate on the English subsets of \textbf{Seed-TTS}
\citep{anastassiou2024seedtts}, \textbf{CV3-eval}, and \textbf{MiniMax},
comprising 1,088, 500, and 100 target utterances, respectively. Within
each benchmark comparison, all systems use the same target texts and
reference recordings. References are processed with RE-USE enhancement
\citep{fu2026reuse}, VAD-based trimming of leading and trailing silence,
and loudness normalization. Speaker similarity (SIM) is measured as the
cosine similarity between WavLM-based speaker embeddings of the
processed reference and generated speech \citep{chen2021wavlm}. Both
waveforms undergo the same loudness normalization before SIM scoring.
Appendix D additionally reports SIM against the original references.

Intelligibility is measured using word error rate (WER), with
transcriptions produced by Whisper large-v3 \citep{radford2023whisper}.
We apply the same text normalization to the target text and ASR
transcription before computing WER. Competing models use their official
inference implementations and default sampling settings. All systems are
scored using the same ASR and speaker-verification models. Appendix A
provides the preprocessing, scoring, and inference parameters.

For reference studies, we use 30 identities from EARS
\citep{richter2024ears}, with four target texts per identity. Similarity
is measured against a different real recording of the same person,
rather than against the reference used to compute \(\mathbf{Z}\).
Reference studies isolate changes in conditioning within each identity.

\section{From Speech Generation to Reference
Encoding}\label{from-speech-generation-to-reference-encoding}

\subsection{Transferring generative capabilities to reference
encoding}\label{transferring-generative-capabilities-to-reference-encoding}

A pretrained TTS generator provides an effective foundation for
reference encoding. Table 1 separates the effects of pretrained
initialization, reference text, and weight sharing, showing that the
gains arise from reusing text--speech modeling capabilities rather than
weight sharing alone.

\textbf{Pretraining transfers useful encoding capabilities.} With the
architecture and training setup fixed, initializing the independent
reference encoder from the TTS backbone raises SIM from 0.400 to 0.530
on Seed-TTS, from 0.342 to 0.469 on CV3-eval, and from 0.391 to 0.592 on
MiniMax. Because this encoder is trained independently of the generator,
the improvement demonstrates that pretrained generative representations
can be repurposed for voice compression without requiring continued
weight sharing.

\textbf{Reference text strengthens this transfer.} Adding the paired
transcript improves SIM on every benchmark for both independent and
shared encoders. The benefit therefore extends beyond a particular
parameter-sharing arrangement: the pretrained backbone becomes a more
effective reference encoder when it receives both what was said and how
it was spoken. Section 6 investigates how this correspondence shapes the
representations used for voice reconstruction.

\textbf{Sharing makes this capability compact.} The independent
text-and-audio encoder achieves the highest SIM, while RSE retains most
of this benefit with 118.2M rather than 228.5M encoding-and-generation
parameters---a 48\% reduction. We therefore adopt sharing to reuse the
backbone's capabilities without maintaining a second Transformer.
Together, these results support text-aware encoding with a pretrained
TTS backbone, with sharing providing a parameter-efficient realization.

We next examine what the learned representation captures, using the same
frozen full RSE-TTS model throughout Sections 5.2--6.

\subsection{Eight vectors can benefit from more
speech}\label{eight-vectors-can-benefit-from-more-speech}

Can eight vectors preserve the additional voice information available in
a longer recording? We study 30 speakers from EARS
\citep{richter2024ears}. For each speaker, we take progressively longer
excerpts from the start of the same recording, approximately 3, 6, 12,
and 24 seconds long. Each excerpt and its corresponding transcript are
encoded into eight vectors, which condition synthesis of the same four
target texts across all reference lengths. The auxiliary 192-dimensional
speaker embedding is always extracted from the 24-second reference and
held fixed. We compute SIM against a separate real recording of the same
speaker. Mean SIM increases from 0.378 to 0.448, 0.518, and 0.569 as the
reference grows longer (Figure 2).

\begin{figure}[t]
\centering
\includegraphics[width=\columnwidth]{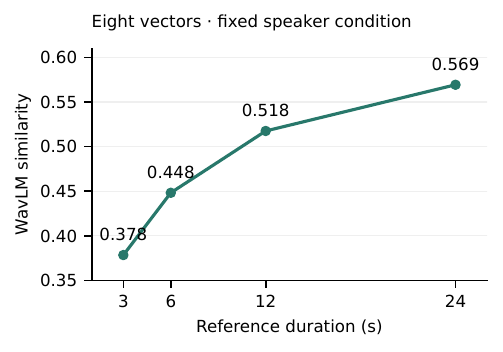}
\caption{Longer reference speech improves similarity even though target
generation always receives eight compressed vectors. Points show
measured means. The auxiliary speaker vector is fixed across durations.}
\end{figure}

Extending the reference from 3 to 12 seconds increases SIM by 0.139
(95\% CI: {[}0.112, 0.165{]}), with gains for 28 of 30 identities. All
30 improve at 24 seconds.

These results show that the eight-vector representation captures
additional speaker information from longer references, improving speaker
similarity without increasing the reference context used during
generation.

\subsection{The vectors carry delivery
information}\label{the-vectors-carry-delivery-information}

Do the eight vectors also preserve how a speaker talks, beyond their
voice identity? We use parallel recordings from EARS in which the same
speaker reads the same passage in regular, fast, slow, and high-pitched
styles. Each recording is encoded with the same reference transcript,
and the resulting eight vectors condition synthesis of the same four new
sentences. The auxiliary speaker embedding is fixed to that extracted
from the speaker's regular reading, so only the reference recording
supplied to the encoder changes.

Compared with conditioning on the regular reading, fast references
produce outputs that are 7.0\% shorter, while slow references produce
outputs that are 13.4\% longer. These durations include pauses.
High-pitched references raise the generated speech's pitch by an average
of 6.01 semitones.

The eight-vector representation transfers speaking-rate and pitch
variations to new sentences even when the auxiliary speaker embedding is
fixed.

\subsection{Complementary information beyond a conventional speaker
embedding}\label{complementary-information-beyond-a-conventional-speaker-embedding}

RSE-TTS conditions generation on both the eight reference vectors and a
conventional 192-dimensional speaker embedding. Do these representations
provide complementary voice information, or does one simply duplicate
the other? We test how well each representation can predict a speaker's
voice characteristics in a different recording.

We construct reference--target pairs for 316 speakers by extending
Seed-TTS references with same-speaker recordings from Common Voice. A
regularized linear predictor estimates the target recording's WavLM
embedding from the conventional speaker embedding, eight reference
vectors, or their concatenation. We evaluate cosine similarity on 63
held-out speakers.

\begin{table}[t]
\centering\small
\setlength{\tabcolsep}{3pt}
\begin{tabular}{@{}lr@{}}
\toprule
Reference representation & Cosine \ensuremath{\uparrow} \\
\midrule
192-dim. speaker embedding & 0.438 \\
Eight reference vectors & 0.439 \\
Both combined & \textbf{0.469} \\
\bottomrule
\end{tabular}
\caption{Predicting the speaker embedding of another real utterance from the same
person. Results are averaged over 63 unseen speakers.}
\end{table}

The two representations perform similarly when used alone, but combining
them improves prediction for 58 of the 63 test speakers. The mean gain
over the conventional embedding is 0.031 (95\% CI: {[}0.025, 0.037{]}).
These results show that the eight vectors and the conventional speaker
embedding provide complementary information for predicting voice
characteristics across utterances. Predictor configurations are
described in Appendix E.

\section{How Reference Text Improves Voice
Encoding}\label{how-reference-text-improves-voice-encoding}

Reference text improves speaker similarity in both pretrained encoder
designs (Table 1). We now investigate how it shapes reference-speech
representations in RSE and whether these changes reach the eight vectors
that condition generation. All experiments in this section use the
frozen full RSE-TTS model and 30 EARS speakers, with four target texts
per speaker. Reference audio, target text, auxiliary speaker embeddings,
and generation seeds are held fixed within each comparison.

\subsection{Disrupting the reference transcript reduces speaker
similarity}\label{disrupting-the-reference-transcript-reduces-speaker-similarity}

We first encode each reference recording twice: once with its correct
transcript and once with the same transcript tokens in reverse order.
Each encoding produces eight vectors, which we use to synthesize the
same new sentences. This changes the order of the reference text without
changing the recording or the set of text tokens.

Mean SIM falls from 0.538 with correct text to 0.497 with reversed text.
Thus, providing the same tokens is not enough: their order matters for
voice reconstruction. Reversal preserves the token inventory while
disrupting textual coherence and transcript--audio correspondence. We
use this perturbation to trace how reference text affects the
representations used for generation.

\subsection{Locating text-dependent changes in reference
encoding}\label{locating-text-dependent-changes-in-reference-encoding}

\textbf{Attention reveals interactions between reference speech and
text.} In layer 3, several heads show clearer sequential audio-to-text
attention patterns with the correct transcript; these patterns weaken or
disperse when the transcript is reversed (Appendix G, Figure 4). These
observations motivate interventions on reference-audio states to test
their influence on the generated voice.

\textbf{SIM gradients identify positions for intervention.} We estimate
the sensitivity of generated speaker similarity to internal encoder
states. The eight query positions show high per-position sensitivity
throughout the encoder, while reference-audio positions remain sensitive
through intermediate layers (Appendix G). These observations motivate
interventions on both groups at layers 3, 6, and 9. We then test whether
replacing their states transfers the benefit of correct reference text
to generation.

\begin{figure*}[t]
\centering
\includegraphics[width=\textwidth]{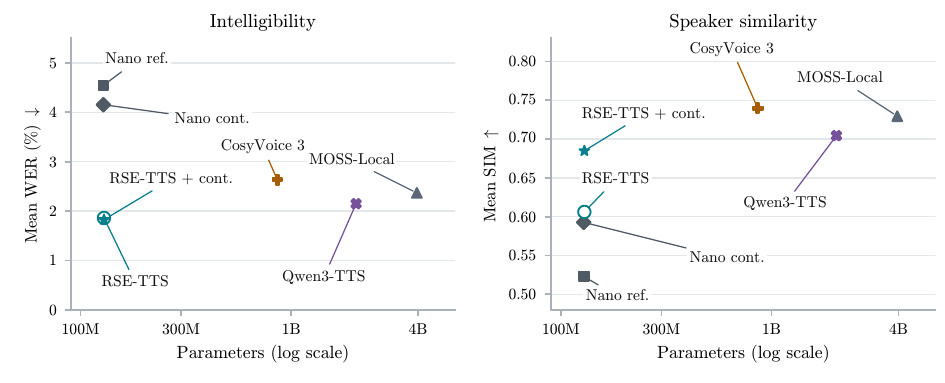}
\caption{Model size versus intelligibility and speaker similarity. WER and SIM
are macro-averaged with equal weight across Seed-TTS, CV3-eval, and
MiniMax. Parameter counts include the generator and active waveform
decoder. Lower WER and higher SIM are better; the horizontal axis uses a
logarithmic scale. ``Ref.'' denotes reference conditioning and ``cont.''
denotes acoustic continuation.}
\end{figure*}

\subsection{Restoring internal representations recovers speaker
similarity}\label{restoring-internal-representations-recovers-speaker-similarity}

Building on the distinction between decodable speaker information and
its causal influence on synthesis \citep{zheng2026decodability}, we test
whether restoring text-dependent reference states recovers the generated
voice. We replace hidden vectors between the correct- and reversed-text
encoding runs using activation patching \citep{zhang2024patching}. After
each replacement, the encoder continues normally, and its final eight
vectors condition ordinary autoregressive speech generation.

For example, to restore layer-3 reference-audio representations, we run
the encoder with reversed text through layer 3, then replace the hidden
vectors at reference-audio positions with those from the correct-text
run. Text and query positions are left unchanged. This tests whether
correcting the representation of the recording can recover voice
similarity even though the transcript remains reversed. We also perform
the opposite replacement: inserting reversed-text representations into
the correct-text run to test whether similarity deteriorates.

\begin{table}[t]
\centering\small
\setlength{\tabcolsep}{3pt}
\begin{tabular}{@{}lrrr@{}}
\toprule
Patched positions & Layer & Restore & Disrupt \\
\midrule
No patch & --- & 0.497 & 0.538 \\
Reference audio & 3 & 0.545 & 0.504 \\
Reference audio & 6 & 0.538 & 0.512 \\
Reference audio & 9 & 0.499 & 0.538 \\
Eight queries & 3 & 0.503 & 0.539 \\
Eight queries & 6 & 0.512 & 0.541 \\
Eight queries & 9 & 0.542 & 0.507 \\
\bottomrule
\end{tabular}
\caption{Speaker similarity after replacing reference-audio or query
representations at layers 3, 6, and 9. Restoration inserts correct-text
representations into the reversed-text run; disruption makes the
opposite replacement. Each condition generates the same 120 targets from
30 speakers.}
\end{table}

\textbf{Correcting reference-audio representations restores the voice.}
Replacing layer-3 audio states raises SIM from 0.497 to 0.545,
recovering the loss caused by transcript reversal. The reverse
replacement lowers SIM from 0.538 to 0.504. Layer-6 audio replacements
show the same direction, while layer-9 audio replacements have little
effect (Table 3). The layer-3 restoration gain is 0.048 (95\% CI:
{[}0.032, 0.065{]}). These results show that text-dependent changes in
reference-audio representations influence the voice subsequently
generated.

\textbf{Useful information is present in the later query
representations.} Replacing the eight query states after layer 9 raises
SIM from 0.497 to 0.542; the reverse replacement lowers it from 0.538 to
0.507. The restoration gain is 0.045 (95\% CI: {[}0.031, 0.060{]}).
Query replacements at earlier layers are less effective.

Random perturbations of matched magnitude do not reproduce these
restorations. The bidirectional layer-3 audio-state effect also
replicates with free-form reference recordings (Appendix E).

The interventions reveal a layer-dependent contrast. At layers 3 and 6,
restoring reference-audio states recovers speaker similarity, whereas
restoring query states has a smaller effect. At layer 9, this pattern
reverses: restoring audio states has little effect, while restoring the
eight query states recovers similarity. Together, these results show
that the benefit of reference text is expressed in intermediate acoustic
representations and subsequently captured in the compact representation
used for generation.

\section{What Can a Compact Generator
Achieve?}\label{what-can-a-compact-generator-achieve}

\textbf{Compact conditioning versus continuation.} A continuation-only
model, trained with the same initialization, data, and schedule but
without reference-encoder or speaker-embedding conditioning, achieves
higher SIM. RSE yields lower WER on all three benchmarks (Table 4),
using fixed-size reference conditioning instead of the full acoustic
history.

\begin{table}[t]
\centering\small
\setlength{\tabcolsep}{3pt}
\begin{tabular}{@{}lrr@{}}
\toprule
Benchmark & Continuation & RSE \\
\midrule
Seed-TTS & 1.45 / \textbf{0.591} & \textbf{1.33} / 0.552 \\
CV3-eval & 3.48 / \textbf{0.526} & \textbf{2.63} / 0.499 \\
MiniMax & 1.77 / \textbf{0.684} & \textbf{0.64} / 0.615 \\
\bottomrule
\end{tabular}
\caption{Matched continuation-only and RSE systems. Cells report WER (\%) \ensuremath{\downarrow} / SIM
\ensuremath{\uparrow} as in Table 1.}
\end{table}

We compare the full RSE-TTS model with the official
MOSS-TTS-Nano,\footnote{https://github.com/OpenMOSS/MOSS-TTS-Nano}
CosyVoice 3,\footnote{https://huggingface.co/FunAudioLLM/Fun-CosyVoice3-0.5B-2512}
MOSS-TTS-Local-Transformer,\footnote{https://github.com/OpenMOSS/MOSS-TTS}
and Qwen3-TTS.\footnote{https://huggingface.co/Qwen/Qwen3-TTS-12Hz-1.7B-Base}
We report compact conditioning and RSE plus continuation. The latter
retains reference speech as acoustic history alongside the reference and
target transcripts, trading a larger context for higher similarity; only
new speech is scored. Figure 3 compares quality and generation-chain
size; Appendices F and B provide benchmark results and parameter counts.

At nearly the same generation-chain size as the official Nano model,
RSE-TTS improves both WER and SIM in the corresponding
reference-conditioning and continuation modes on all three benchmarks.
Combining RSE with acoustic continuation further improves speaker
similarity while maintaining a similar average WER.

RSE-TTS also offers a competitive trade-off against larger systems. With
acoustic continuation, it achieves lower WER than CosyVoice 3 and
MOSS-TTS-Local on each of the three benchmarks, while approaching their
speaker similarity. Its generator and waveform decoder total 129M
parameters---6.6\ensuremath{\times} and 30.5\ensuremath{\times} smaller than those two systems,
respectively.

\section{Conclusion}\label{conclusion}

RSE-TTS reuses a pretrained speech generator for text-aware reference
encoding. Controlled experiments establish the benefits of pretrained
initialization and paired transcripts; sharing the backbone saves 48\%
of encoder--generator parameters. Activation interventions trace the
text benefit through acoustic representations into eight vectors that
preserve voice identity and delivery. The full compact system improves
WER and SIM over official MOSS-TTS-Nano and, with acoustic continuation,
approaches larger models' similarity while achieving lower WER than the
evaluated CosyVoice 3 and MOSS-TTS-Local systems.

\clearpage
\section{Limitations}\label{limitations}

Compact reference conditioning still trails acoustic continuation in
speaker similarity. Diagnostics from an earlier checkpoint show that a
similarity gap is already present near the start of generation and
persists toward the end, while sensitivity to reference conditioning
decreases as acoustic history grows (Appendix C). This pattern suggests
that teacher forcing encourages reliance on acoustic history, allowing
an initial voice mismatch to persist; information loss during
compression is another possible contributor. RSE requires a reference
transcript, making transcription robustness an important consideration
for audio-only applications. Our experiments focus on English.

\section{Ethics and Responsible Use}\label{ethics-and-responsible-use}

Voice-cloning systems can support accessibility, localization, and
personalized speech interfaces, but can also enable impersonation.
Deployment should require appropriate consent and authorization for
reference recordings and clearly communicate the synthetic nature of
generated speech where needed. Any public release should comply with the
licenses and privacy requirements of the underlying data sources.

\bibliography{references}
\clearpage
\appendix
\onecolumn
\nolinenumbers
\makeatletter
\begin{multicols}{2}

\section{Training and Inference
Details}\label{training-and-inference-details}

\textbf{Training loss.} We average token cross-entropy separately for
the native text/control channel, which includes the continuation and EOS
predictions, and for each of the 16 codec channels. The respective
weights are 1 and 2 per codec channel. The final loss is the weighted
sum of these channel means divided by the sum of their active weights
(33 when all channels are active). Only target-branch predictions are
supervised; gradients propagate through the reference pass into the
shared backbone.

\textbf{Controlled ablation.} All five models use the same 20k-hour
subset (approximately 8.6M pairs) and pretrained generator checkpoint,
with fresh optimizers. The random independent reference encoder is
initialized from scratch. The pretrained independent encoders copy
Global and its input embeddings from that checkpoint before training,
then update separately; the shared encoders use the generator's
parameters throughout. The original small speaker encoder is replaced in
every variant. Training seed is 42. The learning rate starts at
\(10^{-6}\), reaches \(2\times10^{-5}\) at 0.05 epoch, stays constant
through 0.5 epoch, then cosine-decays to \(10^{-6}\) at one epoch. All
variants finish at step 27,753.

\textbf{Full RSE-TTS model.} We initialize from a pretrained English TTS
model with a shared-backbone reference encoder and continue training on
the larger corpus. The reference studies and full-system comparisons use
the same frozen model.

\textbf{RSE-TTS sampling.} Audio sampling uses temperature 0.4,
top-\(k\) 10, top-\(p\) 0.8, and repetition penalty 1.2. Text/EOS
sampling uses temperature 1.0, top-\(k\) 50, and top-\(p\) 0.99. The
final benchmark permits at most 400 codec frames and uses no text-length
guard. Controlled ablations use the same length-guard configuration
across all variants. Mechanistic studies use a fixed paired generation
seed, 2026102101, and no length guard.

\textbf{Reference preprocessing versus scoring.} In both benchmark
comparison groups, all models receive the same reference waveform after
RE-USE enhancement \citep{fu2026reuse}, VAD edge trimming with 150 ms
padding while preserving interior pauses, and speech loudness
normalization toward \ensuremath{-}23 LUFS with gain bounded to ±20 dB and a \ensuremath{-}1 dBFS
peak cap. Complete remaining references are used rather than fixed
three-second crops. Primary SIM uses the processed prompt; original,
pre-enhancement reference anchors are scored separately in Appendix D.
Both scoring waveforms are normalized to \ensuremath{-}23 LUFS without additional VAD
or enhancement. ASR uses Whisper large-v3 \citep{radford2023whisper}
through faster-whisper, with int8/float16 execution, English language,
beam size five, temperature zero, no VAD, and no conditioning on
preceding transcription. Text normalization for WER does not change
synthesis inputs. All benchmark ablations use this frontend. The
reference-length, expression, and mechanism studies use the
within-speaker protocols described in Sections 5.2--6, with separate
real recordings as SIM anchors.

\textbf{Aggregation.} System-level WER averages per-utterance normalized
errors divided by reference word count. The ablation uses the sum of
normalized errors divided by the sum of reference words. SIM averages
valid waveform scores; empty generations contribute to WER and are
excluded from SIM. Per-system scoring counts are retained with the
evaluation records. Controlled interventions use identity-cluster
bootstrap intervals, with Holm adjustment within activation-test
families. Complete-system tables report point estimates.

\textbf{Reference-study setup.} Six additional EARS identities are
reserved for development choices. Paired confidence intervals use
identity-level resampling. The fast and high-pitched comparisons include
30 speakers, and the slow comparison includes 28. Additional recordings
come from Common Voice 17.0, matched to Seed-TTS references by
contributor ID. Each of 316 speakers contributes one reference--target
pair of different utterances. Training, development, and test sets
contain 190, 63, and 63 disjoint speakers, respectively. Scores measure
cross-utterance embedding prediction rather than synthesized speech.

\end{multicols}
\clearpage
\noindent\begin{minipage}{\textwidth}
\section{Parameter Accounting}\label{parameter-accounting}
\begingroup
\def\@captype{table}

\centering\small
\fitwidetable{%
\begin{tabular}{lrrr}
\toprule
System & TTS generator & Active waveform decoder & Generation chain \\
\midrule
Official MOSS-TTS-Nano & 117.3M & 11.1M & 128.4M \\
RSE-TTS & 118.2M & 11.1M & \textbf{129.2M} \\
CosyVoice 3 & 506.1M & 353.0M & 859.2M \\
MOSS-TTS-Local-Transformer & 3,060.6M & 887.4M & 3,948.0M \\
Qwen3-TTS-12Hz-1.7B-Base & 1,916.7M & 114.3M & 2,031.0M \\
\bottomrule
\end{tabular}}
\caption{Unique parameter counts for the TTS generator and active waveform
decoder. Tied weights are counted once; reference-preparation modules
are excluded.}

\endgroup\par\smallskip
\end{minipage}\par\medskip
\begin{multicols}{2}

The official Nano generator has 117.3M unique parameters and uses the
same 11.1M codec decode path, totaling 128.4M.

The RSE-TTS generator contains 118.2M unique parameters. Its active
codec decode path has 11.1M parameters; the complete codec, including
the encoder, has 22.0M. Reference preparation additionally uses the
conditioning speaker extractor and VAD. Reference-preparation modules
are excluded from the generation-chain count. The shared Global backbone
is counted once despite its use in two passes. Table 1 separately counts
unique trainable parameters across reference encoding and generation:
118.2M with sharing and 228.5M with an independent reference backbone.
The latter adds Global and input embeddings only, not a second Local
transformer. Once reference vectors are cached, both designs use the
same target-generation architecture.

CosyVoice 3's generator has 506.1M unique parameters. Its flow and
vocoder contribute 332.3M and 20.8M, respectively. The MOSS-TTS-Local
generator has 3.061B parameters; its active decoder and decode-side
quantizer have 887.4M. Qwen3-TTS has 1.917B generator parameters and
114.3M active waveform-decoder parameters; its 12.0M-parameter reference
speaker encoder is listed separately. Reference codec encoders,
evaluation networks, and ASR are excluded from all generation-chain
totals. Buffer tensors and tied aliases are not counted as additional
parameters.

\section{Earlier Onset and History
Diagnostic}\label{earlier-onset-and-history-diagnostic}

We use an earlier shared audio-reference checkpoint to diagnose reliance
on acoustic history, with 18 target texts drawn equally from the three
benchmarks and two generation seeds. Windowed analysis retained 23
paired outputs covering 12 texts. Mean first-two-second and
last-two-second SIM were 0.435 and 0.426 under compressed conditioning,
versus 0.594 and 0.577 under continuation. In fixed-history
interventions, the first-codebook distribution's Jensen--Shannon
divergence after a reference-prefix swap fell from 0.199 in the first
0.4 seconds to 0.015 after four seconds. These diagnostics motivate the
acoustic-history hypothesis discussed in Limitations.

\end{multicols}
\clearpage
\noindent\begin{minipage}{\textwidth}
\section{Original-Reference
Similarity}\label{original-reference-similarity}
\begingroup
\def\@captype{table}

\centering\small
\fitwidetable{%
\begin{tabular}{lrrr}
\toprule
System & Seed-TTS SIM & CV3-eval SIM & MiniMax SIM \\
\midrule
Independent audio (random) & 0.398 & 0.338 & 0.395 \\
Shared audio & 0.495 & 0.432 & 0.552 \\
\textbf{Shared text + audio} & 0.549 & 0.493 & 0.614 \\
Official MOSS-TTS-Nano, reference clone & 0.477 & 0.427 & 0.658 \\
Official MOSS-TTS-Nano, continuation & 0.553 & 0.490 & 0.723 \\
\textbf{RSE-TTS} & 0.595 & 0.535 & 0.674 \\
RSE-TTS + continuation & 0.667 & 0.621 & 0.752 \\
CosyVoice 3, 0.5B-2512, non-RL & 0.702 & 0.672 & 0.825 \\
MOSS-TTS-Local-Transformer, reference clone & 0.702 & 0.651 & 0.812 \\
Qwen3-TTS Base (supplementary) & 0.713 & 0.679 & 0.699 \\
\bottomrule
\end{tabular}}
\caption{Speaker similarity against the original reference recordings. The first
three rows cover the random independent and two shared encoders;
remaining rows compare full systems. Generated audio is unchanged from
the corresponding processed-reference evaluations.}

\endgroup\par\smallskip
\end{minipage}\par\medskip
\begin{multicols}{2}

The main benchmark tables use enhanced reference anchors. Here, we score
the same generated waveforms against the original recordings with the
same loudness normalization; only the reference anchor changes.

\section{Additional Reference-Study
Controls}\label{additional-reference-study-controls}

\textbf{Delivery transfer.} The primary pitch extractor measures a mean
shift of +6.01 semitones; a second extractor measures +4.42 semitones,
with a positive direction for all 30 identities.

\textbf{Embedding probes.} We freeze RSE-TTS and extract its eight
reference vectors before adding the auxiliary speaker projection.
Flattening these vectors gives 6,144 features; the conventional speaker
embedding contributes 192 features, and concatenating both gives 6,336.
The conventional embedding and the 256-dimensional WavLM prediction
target are L2-normalized. Each input block is standardized per
coordinate using training-set statistics, then divided by the square
root of its dimension; the combined predictor concatenates these
normalized blocks.

Each predictor uses ridge regression with an unpenalized intercept,
minimizing mean squared embedding error plus an L2 weight penalty. We
select the penalty coefficient independently for each representation
from powers of ten between \(10^{-6}\) and \(10^{2}\) by development-set
mean cosine similarity. All three select \(10^{-2}\). Predictors are
fitted on the 190 training speakers and evaluated on the 63 held-out
test speakers, without refitting on the development set. Confidence
intervals use 2,000 speaker-level bootstrap resamples.

\textbf{Activation interventions.} With free-form reference content,
correct and reversed transcript baselines are 0.515 and 0.478. Layer-3
reference-audio restoration raises the latter to 0.509, a gain of 0.032
(95\% CI: {[}0.020, 0.045{]}); reverse disruption lowers the correct
condition to 0.479. Norm-matched random controls fail to reproduce the
restoration. A separately selected single-head intervention has no
reliable effect.

\textbf{SIM-gradient approximation and validation.} Discrete
autoregressive sampling prevents direct differentiation of the complete
generation process. We construct a local surrogate by holding sampled
history fixed and using hard-forward, soft-backward codebook
expectations. This straight-through-style surrogate
\citep{bengio2013straightthrough} yields an approximate gradient of
WavLM similarity with respect to the eight-vector prefix.

We perturb the prefix in positive, negative, or random directions with
the same norm, equal to 1\% of the original prefix norm, and then
perform fresh autoregressive generation with paired sampling seeds. Mean
SIM is 0.538 without perturbation, 0.564 along the positive gradient,
0.519 along its negative, and 0.540 for a random direction. Positive
minus negative is 0.045 {[}0.033, 0.058{]}, with the positive direction
better for 29 of 30 identity means.

Fresh generation confirms that the surrogate identifies a
similarity-sensitive direction in the compressed representation. This
\textbf{oracle diagnostic} uses the real speaker anchor and the same
WavLM metric for optimization and evaluation. This supports using the
approximation to identify similarity-sensitive representations; the
activation interventions in Section 6 test their role in the
reference-text effect.

\end{multicols}
\clearpage
\noindent\begin{minipage}{\textwidth}
\section{Per-Benchmark System
Results}\label{per-benchmark-system-results}

\begingroup
\def\@captype{table}

\centering\small
\fitwidetable{%
\begin{tabular}{lrrr}
\toprule
System & \begin{tabular}{@{}c@{}}\textbf{Seed-TTS}\\{\footnotesize WER (\%) $\downarrow$ / SIM $\uparrow$}\end{tabular} & \begin{tabular}{@{}c@{}}\textbf{CV3-eval}\\{\footnotesize WER (\%) $\downarrow$ / SIM $\uparrow$}\end{tabular} & \begin{tabular}{@{}c@{}}\textbf{MiniMax}\\{\footnotesize WER (\%) $\downarrow$ / SIM $\uparrow$}\end{tabular} \\
\midrule
Official MOSS-TTS-Nano, reference clone & 3.96 / 0.480 & 8.10 / 0.431 & 1.58 / 0.659 \\
Official MOSS-TTS-Nano, continuation & 3.13 / 0.556 & 7.57 / 0.492 & 1.77 / 0.730 \\
\textbf{RSE-TTS} & 1.49 / 0.598 & 3.37 / 0.540 & 0.72 / 0.680 \\
RSE-TTS + continuation & 1.66 / 0.670 & 3.07 / 0.627 & 0.74 / 0.758 \\
CosyVoice 3, 0.5B-2512, non-RL & 1.68 / 0.708 & 3.49 / 0.684 & 2.74 / 0.828 \\
MOSS-TTS-Local-Transformer, reference clone & 2.21 / 0.707 & 3.93 / 0.662 & 0.97 / 0.820 \\
Qwen3-TTS-12Hz-1.7B-Base & 1.25 / 0.717 & 4.50 / 0.692 & 0.69 / 0.705 \\
\bottomrule
\end{tabular}}
\caption{English benchmark results under the common evaluation protocol in
Section 4.2. WER is mean per-utterance text-normalized WER in percent;
SIM uses processed references. Figure 3 averages each metric equally
across the three benchmarks.}

\endgroup\par\smallskip

\section{Attention and Gradient
Diagnostics}\label{attention-and-gradient-diagnostics}
\begingroup
\def\@captype{figure}

\centering
\includegraphics[width=\textwidth]{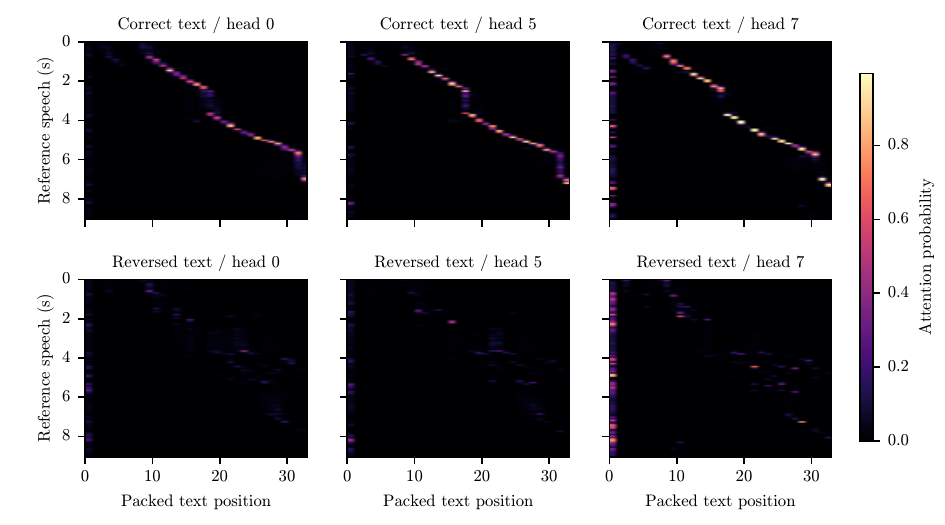}
\caption{Reference-audio attention to text in layer 3 for one fixed EARS speaker.
Each column shows the same head under correct and reversed text, with
identical audio and a shared color scale. Head indices are zero-based.}

\endgroup\par\smallskip
\end{minipage}\par\medskip
\begin{multicols}{2}

Figure 4 illustrates layer-3 reference-audio attention to text for the
first speaker in the evaluation cohort, using the same recording with
correct and reversed transcript tokens. Heads 0, 5, and 7 are the
descriptive examples discussed in Section 6.2; columns retain a common
probability scale. Text positions refer to the actual order supplied to
the encoder in each condition. Audio time is derived from the codec
frame rate, without a fitted alignment shift. The correct transcript
yields clearer sequential correspondence than the reversed transcript.

\end{multicols}
\clearpage
\noindent\begin{minipage}{\textwidth}
\begingroup
\def\@captype{figure}

\centering
\includegraphics[width=\textwidth]{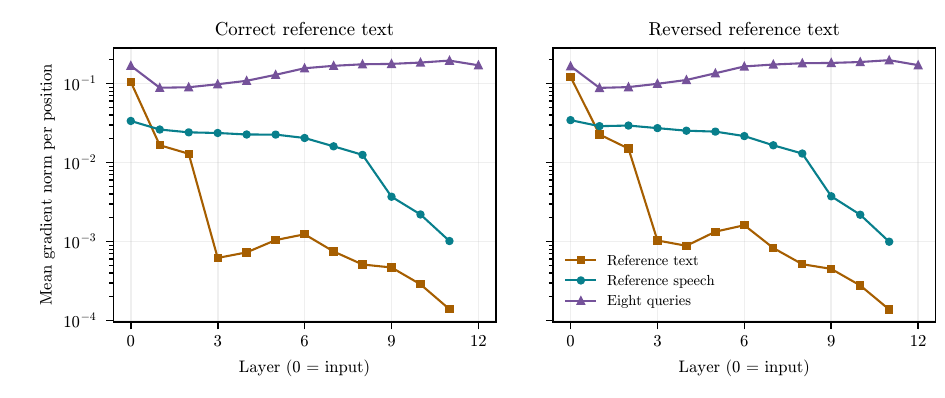}
\caption{Mean per-position gradient norm for reference text, reference speech,
and the eight queries, using the fixed-history soft-backward SIM
estimator at temperature 0.4. The vertical axis is logarithmic. Layer 0
denotes input embeddings. Final-layer text and audio gradients are
exactly zero because only query states are read out; these structural
zeros are omitted from the log plot.}

\endgroup\par\smallskip
\end{minipage}\par\medskip
\begin{multicols}{2}
Figure 5 summarizes approximate SIM-gradient norms over all 120 targets
from 30 speakers. For each example and layer, we average token-gradient
norms within each position group before averaging across examples, so
the longer audio sequence does not dominate merely through its position
count. Query positions have high per-position sensitivity, while
reference-audio sensitivity exceeds text sensitivity through
intermediate layers. Correct and reversed transcripts exhibit similar
aggregate profiles; the functional difference is tested by activation
patching in Section 6.3.

\section{Codec Reconstruction}\label{codec-reconstruction}

We encode and decode the processed reference recordings from all three
English benchmarks using the frozen codec with all 16 codebooks.
References use the same enhancement, silence trimming, and loudness
preprocessing as the system comparison. We compare each reconstruction
with its own input recording using the same WavLM verifier and scoring
normalization. This isolates codec reconstruction from text-to-speech
generation.

\noindent\begin{minipage}{\columnwidth}

\begingroup
\def\@captype{table}

\centering\small
\setlength{\tabcolsep}{3pt}
\begin{tabular}{@{}lrrr@{}}
\toprule
Benchmark & Uses & Unique & SIM \ensuremath{\uparrow} \\
\midrule
Seed-TTS & 1,088 & 666 & 0.894 \\
CV3-eval & 500 & 500 & 0.873 \\
MiniMax & 100 & 2 & 0.948 \\
\bottomrule
\end{tabular}
\caption{Codec reconstruction similarity, averaged over benchmark reference uses.
``Uses'' counts benchmark reference uses; ``Unique'' counts distinct
reference recordings.}

\endgroup\par\smallskip
\end{minipage}

The codec preserves high similarity across all three benchmark reference
sets, supporting our focus on the reference interface and speech
generator. Reconstruction compares the same utterance before and after
coding, whereas voice-cloning evaluation compares different linguistic
content; the two scores answer different questions.

\end{multicols}
\makeatother

\end{document}